\PassOptionsToPackage{table}{xcolor}
\documentclass[sigconf,nonacm]{acmart}
\AtBeginDocument{%
  }

\usepackage{pifont}
\newcommand{\cmark}{\textcolor{green!60!black}{\ding{51}}}
\newcommand{\xmark}{\textcolor{red!70!black}{\ding{55}}}
\newcommand{\dmark}{\textcolor{gray!80}{\textbf{--}}}

\usepackage{array}
\usepackage{multirow}
\usepackage{placeins}
\usepackage{makecell}
\usepackage{tikz}
\usetikzlibrary{positioning,arrows.meta}

\tikzset{
  motif/.style={baseline=-0.6ex},
  mnode/.style={draw, rounded corners=2pt, inner sep=2pt, font=\scriptsize},
  org/.style={mnode},
  conn/.style={mnode},
  medge/.style={-Latex, line width=0.5pt},
  uedge/.style={Latex-Latex, line width=0.5pt}, 
  cedg/.style={-Latex, line width=0.6pt},       
}

\newcommand{\motifTwoHop}{%
\begin{tikzpicture}[motif]
\node[org] (o1) {\texttt{ORG}};
\node[conn, right=10pt of o1] (c) {\texttt{c}};
\node[org, right=10pt of c] (o2) {\texttt{ORG}};
\draw[uedge] (o1) -- (c) node[midway, above, font=\tiny] {$r_1$};
\draw[uedge] (c) -- (o2) node[midway, above, font=\tiny] {$r_2$};
\end{tikzpicture}%
}

\newcommand{\motifThreeHopChain}{%
\begin{tikzpicture}[motif]
\node[org] (o1) {\texttt{ORG}};
\node[conn, right=9pt of o1] (c1) {\texttt{c\textsubscript{1}}};
\node[conn, right=9pt of c1] (c2) {\texttt{c\textsubscript{2}}};
\node[org, right=9pt of c2] (o2) {\texttt{ORG}};
\draw[uedge] (o1) -- (c1) node[midway, above, font=\tiny] {$r_1$};
\draw[uedge] (c1) -- (c2) node[midway, above, font=\tiny] {$r_\text{link}$};
\draw[uedge] (c2) -- (o2) node[midway, above, font=\tiny] {$r_2$};
\end{tikzpicture}%
}

\newcommand{\motifThreeHopBranch}{%
\begin{tikzpicture}[motif]
\node[conn] (c) {\texttt{c}};
\node[org, left=12pt of c] (o2) {\texttt{ORG}};
\node[org, right=12pt of c] (o3) {\texttt{ORG}};
\node[org, above=10pt of c] (o1) {\texttt{ORG}};
\draw[uedge] (o1) -- (c);
\draw[uedge] (o2) -- (c);
\draw[uedge] (c) -- (o3);
\end{tikzpicture}%
}

\begin{document}

\title{FinMultiHopQA: A KG-Grounded Multi-Hop Financial QA Benchmark for Evidence-Driven Reasoning}

\author{Abhinav Arun}
\affiliation{%
  \institution{Domyn}
  \city{New York}
  \country{USA}
}
\email{abhinav.arun@domyn.com}

\author{Reetu Raj Harsh}
\affiliation{%
  \institution{Domyn}
  \country{India}
}
\email{reeturaj.harsh@domyn.com}

\author{Bhaskarjit Sarmah}
\affiliation{%
  \institution{Domyn}
  \city{Gurugram}
  \country{India}
}
\email{bhaskarjit.sarmah@domyn.com}

\author{Stefano Pasquali}
\affiliation{%
  \institution{Domyn}
  \city{New York}
  \country{USA}
}
\email{stefano.pasquali@domyn.com}

\begin{abstract}
Multi-hop financial question answering often requires composing evidence distributed across filing sections, fiscal years, and peer companies. Prior financial QA benchmarks address parts of this problem individually, but none jointly support controlled multi-hop taxonomies, document topologies, motif-level reproducibility, and provenance-linked auditability at scale. We introduce \textbf{FinMultiHopQA}, a benchmark of 15,600 qualitative and quantitative questions, grounded in FinReflectKG, a temporally indexed financial knowledge graph derived from SEC 10-K filings with 17.5M+ triples. The pipeline instantiates \emph{statistically salient} 2--3 hop typed KG motifs across intra-document, inter-year, and cross-company regimes. Beyond closing this gap, the same construction lets us hold each question's underlying evidence constant while varying only its presentation -- embedding it in noisy surrounding context or changing its order -- giving a controlled instrument to study a secondary question at scale: using six such presentations, we test whether reasoning LLMs' performance depends on evidence structure. Results confirm this: despite their reasoning capacity, judged correctness remains consistently lower under less-structured presentations of the same evidence -- most sharply for quantitative cross-company reasoning and 3-hop branch motifs -- indicating that reasoning alone does not fully offset how evidence is structured; KG-linked presentation also uses fewer input tokens by design. We release the benchmark, provenance annotations, model outputs, evaluation records, and supporting technical documentation.
\end{abstract}



\maketitle

\section{Introduction \& Motivation}
\label{sec:intro}

Despite rapid progress in LLM-powered question answering, \emph{reliably} answering analyst-style multi-hop financial queries remains challenging. Evidence needed to resolve a single question is often dispersed across filing sections, fiscal periods, and peer firms, requiring compositional linking of grounded financial facts. Under these conditions, standard retrieval-augmented generation (RAG) pipelines frequently struggle with incomplete or noisy long-context evidence, degrading reasoning reliability.

Existing financial QA benchmarks provide important foundations but leave critical gaps for evaluating analyst-style multi-hop reasoning. Single-document datasets such as FinQA and ConvFinQA~\cite{chen2022finqadatasetnumericalreasoning, chen2022convfinqaexploringchainnumerical} emphasize hybrid table-text numerical reasoning but remain bounded to a single report, limiting cross-company and inter-year composition. Retrieval-oriented suites such as FinanceBench~\cite{islam2023financebenchnewbenchmarkfinancial} stress robustness under noisy contexts yet lack controlled multi-hop regimes or hop-level evidence grounding. General domain multi-hop benchmarks~\cite{yang2018hotpotqa, zhu2024fanoutqa} likewise do not capture finance-specific constraints such as temporally indexed disclosures, peer-relative comparisons, and regulatory semantics.

To address these limitations, we introduce FinMultiHopQA, a large-scale benchmark for multi-hop financial QA grounded in FinReflectKG~\cite{arun2025finreflectkgagenticconstructionevaluation}, a temporally indexed, provenance-linked knowledge graph (KG) derived from SEC 10-K filings; for this benchmark we restrict FinReflectKG to S\&P 100 companies spanning fiscal years 2022--2024. Our generation pipeline combines domain priors conditioned on the Global Industry Classification Standard (GICS) with KG-driven structural control to instantiate analyst-style 2-hop, 3-hop, and branched motifs. Informative connectors are selected using corpus-level Inverse Document Frequency (IDF) specificity to avoid trivial hub connectivity, enabling scalable construction of discriminative, evidence-grounded questions aligned with realistic financial reasoning patterns.

\paragraph{Evidence-sensitivity question.} Every evidence protocol we study (Section~\ref{sec:evidence_protocols}) embeds the \emph{same} ground-truth, answer-bearing chunks; protocols differ only in how that evidence is organized, ordered, and surrounded by noise, never in whether the correct evidence is present. This isolates a single testable question: are reasoning LLMs robust to evidence organization, or does reliable multi-hop reasoning depend on evidence that is already localized and structured before it reaches the model? E1 mimics a graph-RAG-style retrieval system, supplying chunks and triplets aligned to the KG hops required to answer the question; E2--E6 mimic more traditional vector-RAG-style retrieval, varying window size, ordering, and distractor noise around the same underlying evidence. We do not claim to span the breadth of industrial RAG systems, nor is E2--E6 a strongly tuned retriever competing to close this gap; rather, we test whether the absence of KG-derived structure specifically degrades reliability even when the correct evidence remains recoverable in principle. Our core study evaluates open-weight reasoning LLMs, with additional extrapolation to a strong closed-source frontier model and a few finance-performant models (Section~\ref{sec:results_models}).

\textbf{Key contributions:}
\begin{itemize}
    \item \textbf{A large-scale analyst-style multi-hop benchmark.} We construct a 15.6K multi-hop financial QA benchmark grounded in FinReflectKG~\cite{arun2025finreflectkgagenticconstructionevaluation}, spanning intra-document, inter-year, and cross-company reasoning regimes.
    \item \textbf{A domain and structure-aware QA generation pipeline.} We introduce GICS-conditioned connectors coupled with KG-instantiated motif generation that mines statistically informative connectors via IDF specificity~\cite{robertson2009probabilistic}, supporting diverse multi-hop structures and finance-native patterns.
    \item \textbf{Controlled benchmarking of evidence-structure sensitivity.} We define six evidence protocols varying structure, noise, and ordering to test whether reasoning LLMs' performance depends on how evidence is presented; judged correctness is consistently lower under less-structured presentations (a \textbf{16.6--22.9\%} relative gap vs.\ KG-linked provenance), with the same pattern persisting when extrapolated to frontier closed-source and finance-specialized models.
\item \textbf{Public Benchmark Release.} We publicly release the benchmark dataset, including provenance annotations, to support reproducibility and future research.\footnote{Available at: \url{https://huggingface.co/datasets/domyn/FinMultiHopQA}}

\end{itemize}

\section{Related Work}
Multi-hop question answering (QA) has been extensively studied in open-domain settings such as HotpotQA~\cite{yang2018hotpotqa}, 2WikiMultiHopQA~\cite{ho2020constructingmultihopqadataset}, and MuSiQue~\cite{DBLP:journals/corr/abs-2108-00573}, where answering requires composing evidence across multiple documents. Financial QA introduces additional constraints: long, compliance-driven disclosures, hybrid table-text grounding, temporal alignment across fiscal periods, and analyst-centric cross-entity comparisons. These characteristics amplify the difficulty of reliable evidence composition and attribution relative to general-domain multi-hop settings.

\textbf{Single-Document Financial QA.}
FinQA~\cite{chen2022finqadatasetnumericalreasoning} and TAT-QA~\cite{zhu2021tatqaquestionansweringbenchmark} focus on numerical reasoning over single-document table-text contexts, MultiHiertt~\cite{zhao2022multihierttnumericalreasoningmulti} increases structural complexity via hierarchical and interrelated tables, and ConvFinQA~\cite{chen2022convfinqaexploringchainnumerical} extends FinQA into a conversational setting (3{,}892 dialogues, 14{,}115 QA pairs) to improve interaction realism and coreference handling. All remain bounded to single-document evidence and do not enforce cross-filing, cross-period, or cross-company multi-hop composition.

\textbf{Multi-Document and Multi-Hop Financial QA.}
A key limitation of prior financial QA benchmarks is the absence of \emph{controlled} multi-hop regimes spanning cross-company and cross-temporal evidence composition. SEC-QA~\cite{lai2025secqa} broadens multi-document evaluation through refreshable question generation, but does not enforce a unified taxonomy of hop structures or hop-level evidence attribution. Knowledge-graph-driven approaches such as KG-QAGen~\cite{tatarinov2025kgqagenknowledgegraphbasedframeworksystematic} introduce typed motif generation with controllable complexity, yet remain limited for cross-filing temporal and cross-company reasoning. Together, these gaps motivate benchmarks where multi-hop structure and provenance are explicitly controlled, enabling reproducible evaluation and clearer attribution of reasoning failures.

\textbf{Retrieval-Centric and Agentic Evaluation in Finance.}
Financial QA is often bottlenecked by retrieval and evidence localization: FinDER~\cite{choi2025finder} evaluates RAG systems with expert-annotated evidence, DocFinQA~\cite{reddy2024docfinqalongcontextfinancialreasoning} emphasizes evidence localization over long single-company filings, and FinAgentBench~\cite{choi2025finagentbench} studies agentic document navigation -- advancing realistic retrieval evaluation but primarily assessing search behavior rather than structured multi-hop reasoning. Multi-hop queries remain similarly challenging for RAG and agentic information-seeking benchmarks outside finance~\cite{tang2024multihop,xi2025infodeepseek}.

\textbf{Evidence-Grounded and Verifiable Financial QA.}
Complementary benchmarks emphasize auditability and evidence grounding: FinDVer~\cite{zhao2024findverexplainableclaimverification} studies explainable claim verification over long financial documents with expert annotations, and EviFiVQA~\cite{raja2025evifivqa} focuses on evidence-grounded visual QA with fine-grained localization -- strengthening explainability and grounding, but without jointly supporting controlled multi-hop regimes spanning intra-document, inter-year, and cross-company evidence within a unified construction framework.


\textbf{Research Gaps.}
As Table~\ref{tab:kdd_benchmark_comparison} shows, existing financial QA benchmarks do not jointly support controlled multi-hop taxonomies, reproducible motif-level structure, provenance-linked auditability, and large-scale evaluation across structured and unstructured evidence settings. FinMultiHopQA addresses this gap with a unified framework combining KG-controlled motif generation, sector-aware sampling, and provenance-aligned evidence.

\begin{table*}[t]
\centering
\footnotesize
\setlength{\tabcolsep}{2.2pt}
\renewcommand{\arraystretch}{1.10}
\caption{\textbf{Capabilities vs. Verified Open-Source Scale (Financial QA Datasets)} \\
\cmark: explicitly supported as a dataset objective; \xmark: not supported; \dmark: partially supported or not a primary construction target.}
\label{tab:kdd_benchmark_comparison}
\resizebox{\linewidth}{!}{%
\begin{tabular}{l r c c c c c c c c}
\toprule
\textbf{Benchmark} &
\textbf{OS Size} &
\textbf{Multi-hop} &
\textbf{Multi-doc} &
\textbf{Temporal} &
\textbf{Cross-Entity} &
\textbf{Hop-prov.} &
\textbf{Arithmetic} &
\textbf{KG/Motifs} &
\textbf{Reprod. Eval} \\
\midrule

\textit{Gen 1: Arithmetic-centric} & & & & & & & & & \\
FinQA~\cite{chen2022finqadatasetnumericalreasoning} &
8,281 & \xmark & \xmark & \xmark & \xmark & \cmark & \cmark & \xmark & \cmark \\
ConvFinQA~\cite{chen2022convfinqaexploringchainnumerical} &
14,115 & \dmark & \xmark & \xmark & \xmark & \cmark & \cmark & \xmark & \cmark \\
TAT-QA~\cite{zhu2021tatqaquestionansweringbenchmark} &
16,552 & \xmark & \xmark & \xmark & \xmark & \cmark & \cmark & \xmark & \cmark \\
MultiHiertt~\cite{zhao2022multihierttnumericalreasoningmulti} &
10,440 & \dmark & \xmark & \xmark & \xmark & \cmark & \cmark & \xmark & \cmark \\
\midrule
\textit{Gen 2: Retrieval-centric} & & & & & & & & & \\
FinanceBench~\cite{islam2023financebenchnewbenchmarkfinancial} &
\textbf{150}$^{\dagger}$ & \xmark & \dmark & \xmark & \xmark & \cmark & \xmark & \xmark & \dmark \\
FinDER~\cite{choi2025finder} &
5,703 & \dmark & \cmark & \dmark & \dmark & \cmark & \xmark & \xmark & \cmark \\
DocFinQA~\cite{reddy2024docfinqalongcontextfinancialreasoning} &
7,437 & \cmark & \xmark & \xmark & \xmark & \cmark & \cmark & \xmark & \cmark \\

\midrule
\textit{Gen 3: Structural / Agentic} & & & & & & & & & \\
FinAgentBench~\cite{choi2025finagentbench} &
26,000 & \dmark & \cmark & \xmark & \dmark & \cmark & \xmark & \xmark & \cmark \\
SEC-QA~\cite{lai2025secqa} &
-- & \cmark & \cmark & \dmark & \dmark & \dmark & \xmark & \xmark & \dmark \\

\midrule
\rowcolor{green!6}
\textbf{FinMultiHopQA} &
\textbf{15,600} $^{\ddagger}$ & \cmark & \cmark & \cmark & \cmark & \cmark & \cmark & \cmark & \cmark \\
\bottomrule
\multicolumn{10}{l}{$^{\dagger}$ FinanceBench reports 10,231 QA pairs, but releases 150 for open-source evaluation.}\\
\multicolumn{10}{l}{$^{\ddagger}$ FinMultiHopQA contains 36,668 QA pairs in total; 15,600 are publicly released as the standardized evaluation benchmark reported here.}
\end{tabular}%
}

\vspace{2pt}
{\footnotesize
\textbf{Legend:}
OS Size denotes the number of publicly released QA instances. Multi-hop here means composing $\geq 2$ independent evidence facts.
Multi-doc requires evidence across $\geq 2$ filings/documents.
Temporal denotes inter-year reasoning.
Cross-Entity denotes cross-company reasoning.
Hop-prov.\ indicates hop$\rightarrow$evidence alignment.
Reprod.\ Eval indicates the benchmark protocol is reproducible using publicly released evaluation splits/examples.
}
\end{table*}

\section{Dataset Design \& Construction}
\label{sec:dataset_construction}

\subsection{Base Open Source Financial Knowledge Graph: FinReflectKG}
FinReflectKG~\cite{arun2025finreflectkgagenticconstructionevaluation} is a temporally indexed, source-attributed financial knowledge graph derived from SEC 10-K filings. Each extracted triple is linked to its originating document span (file, page, and chunk identifiers), enabling auditable provenance and temporal grounding. \emph{For this work, we restrict FinReflectKG to the S\&P 100 firms spanning fiscal years 2022--2024}. This subset preserves structured firm-level entities and relations while focusing evaluation on recent disclosures. The resulting graph serves as a substrate for controlled multi-hop question generation supporting cross-entity and cross-period reasoning grounded directly in source evidence.
The benchmark does not require this KG to be a complete representation of the filings. Instead, we use the source-linked structure to select and align evidence hops, keeping only instances that are verifiably grounded and genuinely multi-hop.

\subsection{Analyst-Focused Multi-Hop Regimes}
\label{sec:evidence_scopes}

FinMultiHopQA is designed to mirror analyst workflows that compose evidence across filings under three evidence regimes:
\begin{itemize}
    \item \textbf{Intra-document:} same issuer and year, with evidence drawn from distinct sections/pages/facts.
    \item \textbf{Inter-year:} same issuer across different fiscal years.
    \item \textbf{Cross-company:} different issuers within the same GICS sector/sub-industry and same year.
\end{itemize}


\subsection{GICS-Conditioned Industry Grouping and Connector Type Selection}
\label{sec:gics_conditioned}

Analyst workflows typically compare firms within the same industry rather than arbitrary pairs (e.g., NVDA vs.\ AMD, WMT vs.\ TGT). To reflect this realism, we first partition issuers using GICS, ensuring that cross-company motif instantiations occur among economically comparable firms. Within each GICS partition, we constrain the types of intermediate connector entities that govern how multi-hop questions are formed.

\textit{Connector regimes.}
Qualitative questions draw connectors from sector-aligned entity classes (e.g., competitors, risk factors, macro conditions, raw materials, regulatory drivers), suppressing dominance by generic hubs while preserving interpretability. Quantitative questions restrict connectors to \texttt{FIN\_METRIC} etc., enabling grounded arithmetic and cross-entity comparison while retaining industry-aware peer selection.

\textit{Data-driven connector discovery.}
Connector relevance is inferred directly from KG structure. For each
  issuer-connector pair $(o,x)$, let $s=\mathrm{sector}(o)$ and
  $t=\mathrm{type}(x)$, with $N_{s,t}$ the number of issuers in sector $s$
  connected to at least one connector of type $t$ (total $N$).
The joint and marginal distributions are
\[
P(s,t)=\frac{N_{s,t}}{N}, \quad P(s)=\sum_t P(s,t), \quad P(t)=\sum_s P(s,t).
\]
Sector–connector dependence is measured by mutual information:
\begin{equation}
I(S;T)=\sum_{s,t}P(s,t)\log\frac{P(s,t)}{P(s)P(t)}.
\end{equation}
Connector types are ranked by their contribution to $I(S;T)$, identifying those disproportionately associated with a sector relative to global connectivity; node-level informativeness scoring then selects sector-aware connectors types for motif instantiation.

\subsection{Informative Connector Mining}
\label{sec:connector_scoring}

Large-scale motif mining must avoid trivial connectivity dominated by ubiquitous hubs while preserving analyst-relevant structure. Connectors are ranked by IDF specificity -- $\mathrm{idf}(x) = \log(|\mathcal{O}|/\mathrm{df}(x))$ -- which downweights ubiquitous hubs in favour of discriminative, analyst-relevant connectors~\cite{robertson2009probabilistic,salton1988term}, with a minimum support filter $\mathrm{df}(x)\geq\tau$ to avoid noisy signals.

\subsection{Multi-Hop Typed Motif Families}
\label{sec:typed_motifs}

We generate multi-hop QA pairs by instantiating a library of \textbf{typed graph motifs} over FinReflectKG under the controlled regimes of Section~\ref{sec:evidence_scopes}, where each question requires evidence composition across multiple KG edges with explicit provenance constraints (issuer, fiscal year, and page/chunk attribution).

\paragraph{2-hop bridge motifs.}
\label{sec:2hop_bridge_motifs}

We instantiate 2-hop motifs as a bridge pattern between two ORG anchors through a connector node:
\begin{center}
\motifTwoHop
\end{center}
where $c$ is a connector entity (e.g., \texttt{FIN\_METRIC}, \texttt{RISK\_FACTOR}, \texttt{PRODUCT}, \texttt{RAW\_MATERIAL}, \texttt{SEGMENT}, \texttt{EVENT}). This motif yields analyst-style comparison questions such as “how do two issuers relate to the same risk driver or metric disclosure?”

\noindent\textbf{Regime constraints.}
Cross-company enforces $\texttt{ORG}_1\neq\texttt{ORG}_2$ with matched years; inter-year enforces the same ORG with $\texttt{year}(r_1)\neq\texttt{year}(r_2)$; intra-document enforces the same ORG and document with $\texttt{page/chunk}(r_1)\neq\texttt{page/chunk}(r_2)$ (or, when synthesis is required, distinct facts within a shared chunk).

\paragraph{3-hop motifs.}
\label{sec:3hop_motifs}

To capture multi-step reasoning workflows, we instantiate two families of 3-hop motifs.

\subparagraph{Type 1: Connector-chain (two connectors).}
We construct 3-hop chains with two connector nodes linked together, anchored by organizations at both ends:
\begin{center}
\motifThreeHopChain
\end{center}
This family supports questions where the reasoning path passes through a connector--connector association (e.g., \texttt{RAW\_MATERIAL} $\leftrightarrow$ \texttt{EVENT}, or \texttt{RISK\_FACTOR} $\leftrightarrow$ \texttt{FIN\_METRIC}). To encourage diversity, we select connector pairs $(c_1,c_2)$ using IDF ranking within GICS strata and retain top-$K$ pairs per sector/sub-industry.

\noindent\textbf{Regime constraints.}
Constraints mirror the 2-hop case; $r_\text{link}$ must share provenance with ORG-anchor edges: matched years for cross-company and inter-year chains, and $\texttt{doc}(r_\text{link})=\texttt{doc}(r_1)=\texttt{doc}(r_2)$ with distinct pages/chunks for intra-document chains.

\begin{figure*}[t]
\centering
\includegraphics[width=\textwidth]{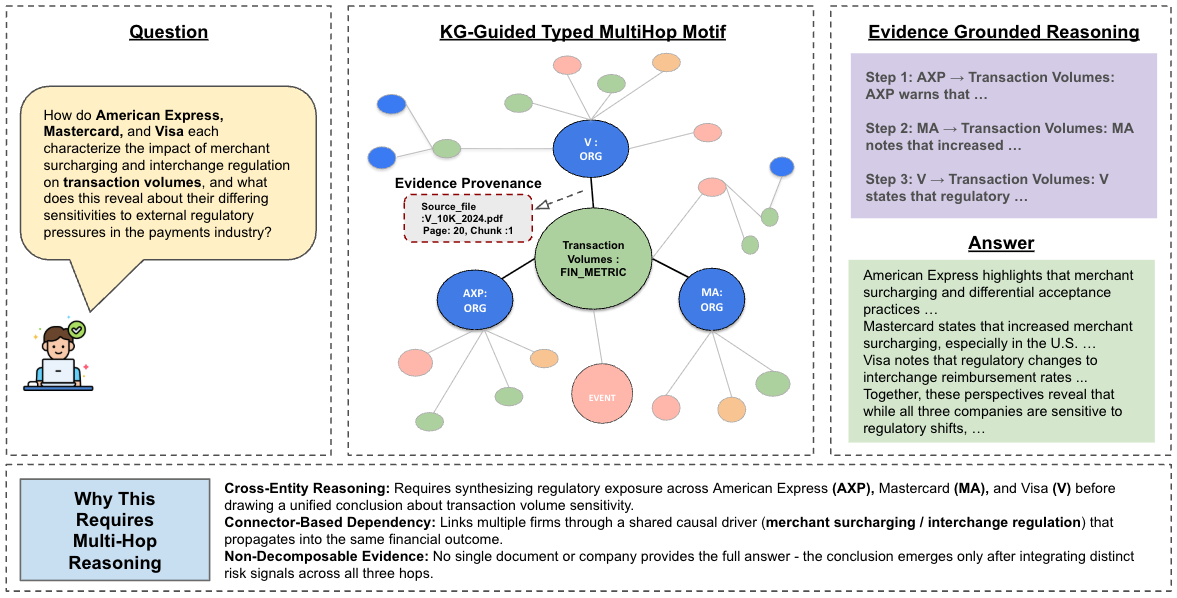}
\Description{Diagram showing three organization nodes (AXP, MA, V) connected through a shared financial connector node, illustrating a cross-company multi-hop motif.}
\caption{
\textbf{Cross-company multi-hop QA example constructed using KG-guided typed motifs.}
The question requires synthesizing regulatory exposure across multiple organizations (AXP, MA, and V) via a shared financial connector.
}
\label{fig:cross_company_multihop_example}
\end{figure*}

\subparagraph{Type 2: Multi-branch (three ORG anchors).}
We also instantiate a bounded branched motif that captures “triangulation” workflows common in analyst research, example shown in Figure ~\ref{fig:cross_company_multihop_example}.
\begin{center}
\motifThreeHopBranch
\end{center}
This motif enforces multi-entity composition while keeping evidence size bounded, and produces questions involving peer grouping, shared exposures, disclosures across three anchors.

\noindent\textbf{Regime constraints.}
Each regime fixes which anchors vary around the shared connector $c$: cross-company requires three distinct ORG anchors within the same year; intra-document requires all three anchor edges to resolve within the same document across distinct pages/chunks (or distinct facts within a shared chunk); inter-year requires one organization spanning three distinct fiscal years.

\subsection{KG-Driven Evidence Extraction and Grounding}
For each motif match, we retrieve provenance-linked evidence chunks corresponding to each hop. Each QA example therefore includes a \emph{structured evidence core}: a set of chunks with explicit hop-to-evidence mapping, enabling controlled evaluation of compositional reasoning under various evidence settings (Section~\ref{sec:evidence_protocols}).

\subsection{QA Generation and Validation}
Natural language QA pairs are generated from KG-motif grounded chunks using a strong reasoning model (Qwen3-235B-A22B), conditioned on motif structure, entities, provenance evidence, and regime constraints.
Both qualitative and quantitative questions are produced with sector-stratified motif sampling.
Every generated pair is then automatically scored, gated, and filtered to the 15{,}600-pair evaluation benchmark through the multi-tier validation procedure detailed in Section~\ref{sec:benchmark_construction}.

\subsection{Controlled Evidence Protocols}
\label{sec:evidence_protocols}

We then evaluate reasoning robustness for different LLMs under six evidence protocols that vary structure, noise, and order effects. "Page" and "chunk" both denote spans of source filing text: provenance is tracked at chunk granularity during construction (Section~\ref{sec:dataset_construction}), while E2--E6 assemble context at page granularity:
\begin{itemize}
    \item \textbf{E1 (KG-linked):} Provenance-linked evidence chunks corresponding to the instantiated KG hops. It is a structured reference point for isolating evidence organization, not an oracle answer condition or idealized ground truth.
    \item \textbf{E2 (Page-window):} A 5-page window per hop (2 pages before and after the provenance page), yielding up to 10 page-window entries for 2-hop and 15 for 3-hop questions; entries can repeat across overlapping hops and may be fewer when the window spills past the document's start or end.
    \item \textbf{E3 (Page-window + distractors):} E2 provenance page(s) augmented with additional irrelevant pages, selected from at least 10 pages away from any target, to simulate retrieval noise and spurious matches.
    \item \textbf{E4 (Relevance-sorted, descending):} E2 context sorted by semantic similarity to the question (most relevant first), testing order sensitivity.
    \item \textbf{E5 (Relevance-sorted, ascending):} E2 context sorted by semantic similarity (least relevant first), an adversarial test.
    \item \textbf{E6 (Semantic retrieval + random order):} Top-$5$ semantically similar pages/chunks retrieved from the same document as the provenance evidence, followed by random shuffling to disentangle relevance from ordering effects.
\end{itemize}

\section{Dataset Statistics and Benchmark Composition}
\label{sec:dataset_stats}

\subsection{Full QA Corpus Statistics}
\label{sec:full_corpus_stats}
The full FinMultiHopQA corpus contains 36,668 multi-hop QA pairs generated over a temporally indexed financial knowledge graph derived from SEC 10-K filings. The dataset spans both qualitative and quantitative reasoning and systematically covers 2-hop and 3-hop dependency structures. The corpus is approximately balanced between qualitative (47.9\%) and quantitative (52.1\%) questions, while 3-hop questions form the majority (67.2\%) of instances (Table~\ref{tab:full_vs_filtered_corpus_stats}). This reflects an intentional emphasis on non-trivial cases requiring aggregation across entities, fiscal periods, or document sections.

\subsection{Evaluation Benchmark Construction}
\label{sec:benchmark_construction}

\begin{table}[t]
\centering
\small
\caption{Hop-count distribution for the full QA dataset and filtered evaluation benchmark.}
\label{tab:hop_distribution_combined}
\begin{tabular}{l|cc|cc}
\toprule
& \multicolumn{2}{c|}{\textbf{Full Dataset}}
& \multicolumn{2}{c}{\textbf{Eval Benchmark}} \\
\textbf{Question Type} & 2-hop & 3-hop & 2-hop & 3-hop \\
\midrule
Qualitative  & 3,934  & 13,624 & 3,000 & 4,800 \\
Quantitative & 8,098  & 11,012 & 3,000 & 4,800 \\
\midrule
Total        & 12,032 & 24,636 & 6,000 & 9,600 \\
\bottomrule
\end{tabular}
\label{tab:full_vs_filtered_corpus_stats}
\end{table}

The final 15,600-pair evaluation benchmark (Table~\ref{tab:full_vs_filtered_corpus_stats}) is assembled from the full 36,668-pair corpus through two automated stages followed by expert validation.

\paragraph{Automated screening and selection.}
Every generated QA pair is scored by an LLM validator on five dimensions (0--10 each, max 50): multi-chunk dependency, financial context, answer grounding, connection grounding, and natural analyst quality.
Hard rejection criteria embedded in the rubric immediately discard pairs that mention document structure, require external knowledge, or assert a fabricated causal link unsupported by evidence. Multi-hop necessity is enforced at three levels (barring some same-chunk intra-document synthesis): (i) \emph{single-chunk non-answerability}, rejecting pairs answerable from any one evidence chunk alone; (ii) \emph{unused-chunk rejection}, requiring every provided chunk to contribute essential facts rather than being superficially mentioned; and (iii) \emph{all-hop necessity}, requiring that removing any single chunk render the question unanswerable or incomplete.
Pairs scoring below 45/50 are rejected or regenerated.
From the passing pool, within each of the 18 question families (two question types multiplied by three 2-hop patterns plus six 3-hop pattern/topology combinations), candidates are ranked by rubric score and connector IDF, and top-ranked examples are retained to form the 15,600-pair benchmark, preserving qualitative/quantitative balance (7,800 each) while emphasizing 3-hop questions (9,600 vs.\ 6,000 two-hop).
\emph{Unless otherwise noted, all experiments, results and distribution plots refer to this filtered benchmark.}

\paragraph{Independent SME quality audit.}
Building on this automated filtering, an independent audit of the resulting benchmark by two finance-domain SMEs found consistently high QA quality: pass rates were 99.7--100\% for multi-hop dependency, 100\% for financial complexity, 98.9--99.4\% for evidence-based answerability, and 100\% for overall dataset relevance. Both SMEs evaluated the same fixed-seed, topology-stratified random sample of 360 pairs (20 from each of the 18 partitions) using these four binary pass/fail criteria, with 353 pairs (98.1\%) passing all four according to both reviewers. For intra-document pairs, the SMEs explicitly treated distinct facts or passages within the same extracted chunk as valid hops when their synthesis was required. The few flags primarily reflected localized answer-generation issues rather than systematic structural defects: the question, evidence, and most of the answer remained consistent, but an occasional detail was unsupported due to forecast-versus-actual or metric conflation, cross-hop arithmetic errors, or entity-scope mismatches. These low-frequency findings affirm the benchmark's overall robustness while motivating targeted checks for calculation accuracy, temporal and metric consistency, and evidence--answer alignment.

\paragraph{Coverage and balance.} Cross-company and inter-year cases are intentionally prominent, increasing evidence fragmentation across entities and reporting periods, while topology ratios are preserved across 2-hop and 3-hop questions -- isolating hop-count effects from document structure bias -- with 3-hop instances further balanced across Type~1 and Type~2 motif families. Each QA pair is further classified into a single, most-relevant financial reasoning category by a rule set over its typed KG schema and motif pattern (e.g., an \texttt{ORG}$\to$\texttt{FIN\_METRIC} edge maps to \emph{Metric Disclosure}); this post-hoc analysis reveals that the connector-chain construction approach naturally yields broad coverage without explicit enforcement: across 15,600 items and 1,919 unique \textsc{(source-node, relation, target-node)} motif patterns, the benchmark spans 13 distinct categories, from Metric Disclosure to the rarer Risk-to-Metric pattern (Table~\ref{tab:category_dist}). Separately, the benchmark also spans all 11 GICS sectors present in the underlying S\&P 100 subset, from Information Technology (14.9\%) to Materials (1.1\%) (Table~\ref{tab:sector_dist}), reflecting the sector-aware connector selection of Section~\ref{sec:gics_conditioned} rather than post-hoc balancing.

\begin{table}[t]
\centering

\small
\caption{Financial reasoning category coverage.}
\label{tab:category_dist}
\resizebox{\linewidth}{!}{%
\begin{tabular}{lr|lr}
\toprule
\textbf{Category} & \textbf{Count (\%)} & \textbf{Category} & \textbf{Count (\%)} \\
\midrule
Metric Disclosure & 4,529 (29.0\%) & Regulatory Compliance & 916 (5.9\%) \\
Debt \& Capital Markets & 1,671 (10.7\%) & Other & 849 (5.4\%) \\
Supply Chain / Dependency & 1,322 (8.5\%) & Investment / M\&A & 589 (3.8\%) \\
Temporal Trend & 1,218 (7.8\%) & Risk Disclosure & 500 (3.2\%) \\
Product / Revenue & 1,179 (7.6\%) & Geographic Operations & 342 (2.2\%) \\
Accounting Policy & 1,175 (7.5\%) & Macro Sensitivity & 125 (0.8\%) \\
Segment Performance & 1,122 (7.2\%) & Risk-to-Metric & 63 (0.4\%) \\
\bottomrule
\end{tabular}%
}
\end{table}

\begin{table}[t]
\centering
\small
\caption{GICS sector coverage.}
\label{tab:sector_dist}
\resizebox{\linewidth}{!}{%
\begin{tabular}{lr|lr}
\toprule
\textbf{Sector} & \textbf{Count (\%)} & \textbf{Sector} & \textbf{Count (\%)} \\
\midrule
Information Technology & 2,321 (14.9\%) & Communication Services & 1,518 (9.7\%) \\
Industrials & 2,106 (13.5\%) & Energy & 853 (5.5\%) \\
Health Care & 2,019 (12.9\%) & Utilities & 853 (5.5\%) \\
Financials & 1,952 (12.5\%) & Real Estate & 371 (2.4\%) \\
Consumer Staples & 1,828 (11.7\%) & Materials & 168 (1.1\%) \\
Consumer Discretionary & 1,611 (10.3\%) & & \\
\bottomrule
\end{tabular}%
}
\end{table}

\section{Experimental Setup}

We evaluate reasoning models' sensitivity to evidence presentation on this multi-hop QA benchmark under six controlled evidence protocols (E1--E6) defined in Section~\ref{sec:evidence_protocols}. Experiments primarily benchmark five open-weight reasoning LLMs : GPT-OSS-120B, GPT-OSS-20B, Nemotron Nano-30B, Qwen3-8B, and Qwen3-32B using identical prompts, decoding settings, and context limits to isolate evidence effects. Semantic retrieval protocol (E6) employs similarity scoring \& mean pooling using the nvidia/nv-embedqa-e5-v5 embedding model. Outputs are evaluated on a 0--10 ordinal correctness scale using two independent LLM judges (Qwen3-235B and Claude Haiku 4.5) \cite{zheng2023judging, gu2025surveyllmasajudge}. Inter-judge reliability is measured with quadratic-weighted Cohen’s $\kappa$ \cite{cohen1968weighted, landis1977measurement} and Spearman correlation \cite{spearman1904proof}, and statistical significance across evidence settings is tested using paired Wilcoxon signed-rank tests\cite{wilcoxon1945individual} with Holm--Bonferroni correction \cite{holm1979simple}.




\section{Experimental Results \& Analysis}
\label{sec:results}

\subsection{Effect of Evidence Settings on Correctness Score \& Token Consumption}
\label{sec:results_evidence}

\begin{figure}[t]
\centering
\includegraphics[width=\linewidth]{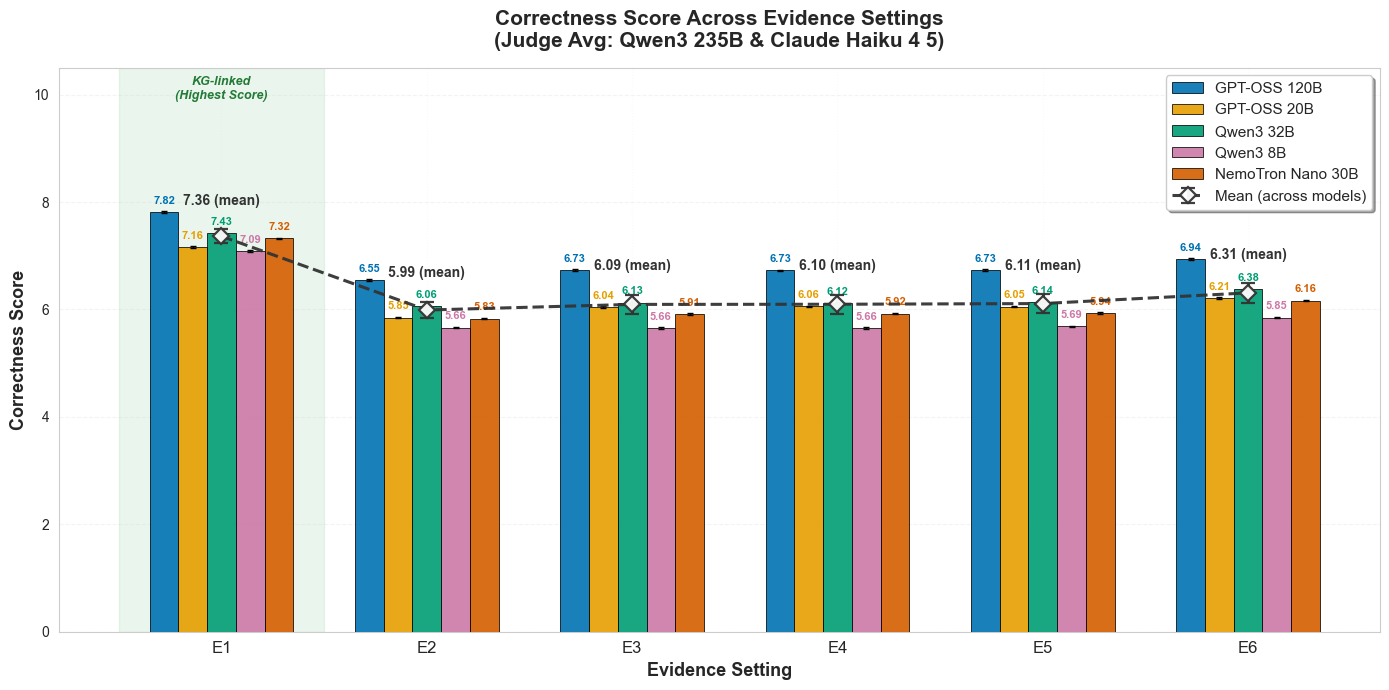}
\Description{Bar chart of average correctness scores for five reasoning models across evidence protocols E1 through E6, with E1 (KG-linked) consistently highest.}
\caption{Average correctness scores across evidence protocols (E1--E6), aggregated over multiple LLM judges.}
\label{fig:multi_judge_average_correctness}
\end{figure}

\begin{figure}[t]
\centering
\includegraphics[width=\linewidth]{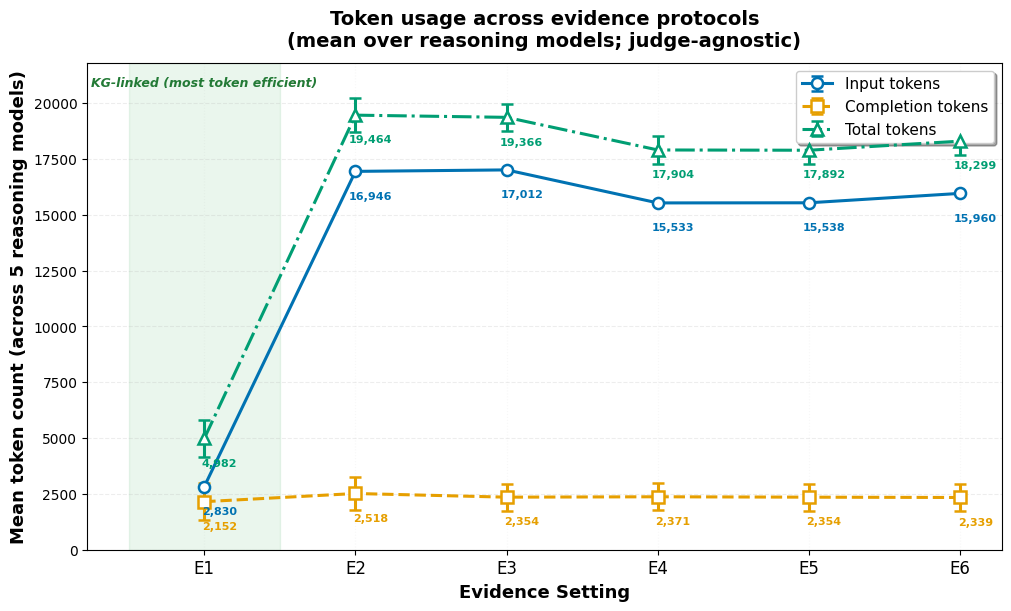}
\Description{Bar chart comparing total token usage across evidence protocols E1 through E6, showing E1 using substantially fewer tokens than E2--E6.}
\caption{Token usage across evidence protocols (mean over reasoning models; judge-agnostic).}
\label{fig:token_usage_evidence_protocols}
\end{figure}

To reduce redundancy across evaluators, we report judge-averaged correctness in Figure~\ref{fig:multi_judge_average_correctness} and judge-agnostic token usage averaged over the 5 reasoning models in Figure~\ref{fig:token_usage_evidence_protocols}.

\paragraph{Key result: KG-linked provenance maximizes correctness.}
Across models, KG-linked provenance evidence (E1) yields the highest average correctness ($7.36$ mean across models) while being the most token-efficient: $4{,}982$ total tokens on average (Figure~\ref{fig:token_usage_evidence_protocols}).
Relative to text-window and semantic-retrieval baselines (E2--E6), E1 improves correctness by
\textbf{+22.9\%} vs.\ E2,
\textbf{+20.9\%} vs.\ E3,
\textbf{+20.7\%} vs.\ E4,
\textbf{+20.5\%} vs.\ E5,
and \textbf{+16.6\%} vs.\ E6 (computed using mean scores in Figure~\ref{fig:multi_judge_average_correctness});
equivalently, E1 is \textbf{+20.3\%} higher than the average of E2--E6.
These gains occur alongside a \textbf{72--74\% reduction} in total token usage relative to E2--E6 ($4{,}982$ vs.\ $17{,}892$--$19{,}464$). This token count difference is driven mainly by the protocol-designed input context: E1 averages $2{,}830$ input tokens vs.\ ${\sim}15.5$--$17$k for E2--E6. Completion usage is also lower under E1 by approximately 10\% ($2{,}152$ vs.\ ${\sim}2{,}387$ tokens on average).

\paragraph{Comparing unstructured baselines: limited gains from ordering and retrieval heuristics.}
Among non-KG provenance protocols, ordering (E4/E5) and distractors (E3) yield only marginal, near-identical recovery over the raw page-window baseline (E2), indicating weak sensitivity to context arrangement once evidence is embedded in noisy material. Semantic retrieval with randomization (E6) is the strongest non-provenance option but remains substantially below E1 while incurring high input token cost (Figure~\ref{fig:multi_judge_average_correctness}).

\subsection{Evaluation on Frontier Closed-Source and Finance-Specialized LLMs}
\label{sec:results_models}

To test whether the E1 advantage persists beyond the five open-weight models of the core study, we evaluate three additional systems selected from the top of the Vals~AI \emph{Finance Agent Benchmark} (v1.1) leaderboard \cite{bigeard2025financeagentbenchmarkbenchmarking}: a leading closed-source frontier model (GPT-5.2), the strongest open-weight system on that leaderboard (DeepSeek-V4, \texttt{think-high}), and a compact finance-competitive model (Gemma~4~31B~IT). GPT-5.2 and DeepSeek-V4 are evaluated under all six evidence protocols (E1--E6) with the Qwen3-235B judge; Gemma~4~31B~IT is evaluated under E1, E2, and E6 only, a deliberate compute-saving choice, since E3--E5 cluster near E2 across every model in the core study (Section~\ref{sec:results_evidence}) and are not expected to alter the qualitative conclusion. GPT-5.2 attains the highest absolute correctness of any model in our study (E1~$=8.65$), exceeding the best open-weight reasoning model (GPT-OSS-120B, E1~$=8.31$), yet still exhibits the same provenance ordering: E1 outperforms the mean of E2--E6 by \textbf{+12.6\%}. DeepSeek-V4 (E1~$=7.95$) and Gemma~4~31B~IT (E1~$=7.80$) follow the same pattern, with E1 lifts of \textbf{+20.3\%} and \textbf{+18.5\%}, respectively. Both within these open-weight models and across the additional closed-source and finance-specialized systems, larger and stronger models achieve higher absolute correctness, but the relative ordering of evidence protocols never changes -- even the strongest frontier model still loses roughly a full point of correctness when KG-linked provenance (E1) is replaced by windowed or retrieved context (E2--E6).

\subsection{Performance by Structural Dimension}
\label{sec:results_docrel}

Qualitative questions outperform quantitative ones in every topology, averaged across all evidence protocols -- most sharply for cross-company reasoning, where precise numerical values, units, and metric identities must be reconciled across independent filings, and least so for inter-year settings, where single-company context partially compensates.
Intra-document questions achieve the highest average in the benchmark, while quantitative cross-company remains the lowest and hardest combination overall.
By motif type, 3-hop branch motifs are the most sensitive to evidence structure ($\Delta_{\text{E1-E2}}=1.63$), as three independent source streams must converge at a shared connector; quantitative branch questions show particularly large degradation under unstructured retrieval.
The E1$\to$E2 gap itself varies by document topology: it is smallest for intra-document questions ($\Delta$ 0.71--0.89 across models), largest for cross-company reasoning ($\Delta$ up to 2.01), and intermediate for inter-year comparisons ($\Delta$ 1.28--1.50), indicating that structured evidence matters most precisely where entity reconciliation must span independent filings.
Across the 13 financial reasoning categories identified in Section~\ref{sec:benchmark_construction}, the E1$-$E2 gap is largest for categories whose evidence is most often scattered across non-adjacent pages -- Metric Disclosure (+1.70), Debt \& Capital Markets (+1.66), and Temporal Trend (+1.62), showing that reasoning over KG-linked provenance remains reliable even for semantically complex financial concepts.

\subsection{Failure Mode Analysis}
\label{sec:failure_modes}

Error analysis corroborates the evidence-structure effect from a different angle: as evidence degrades from E1 to E6, hallucination rises from 13.0\% to 23--26\%, multi-hop failures from 4.6\% to 7.5--8.2\%, and numerical errors from 7.8\% to 12--13\%, corresponding to a failure rate of 30.1\% under E1 versus 48--53\% under unstructured protocols.
These are not merely less-precise answers but qualitatively distinct errors -- hallucinated values, metric substitutions, year transpositions, and multi-hop chain truncation -- rather than uniformly reduced precision.
Residual errors under E1 concentrate in hallucination and numerical precision, consistent with a shift: once the KG resolves entity identity and temporal attribution, arithmetic remains the primary remaining source of error.

\subsection{Evaluation Reliability}
\label{sec:results_agreement}

\begin{figure}[t]
\centering
\includegraphics[width=\linewidth]{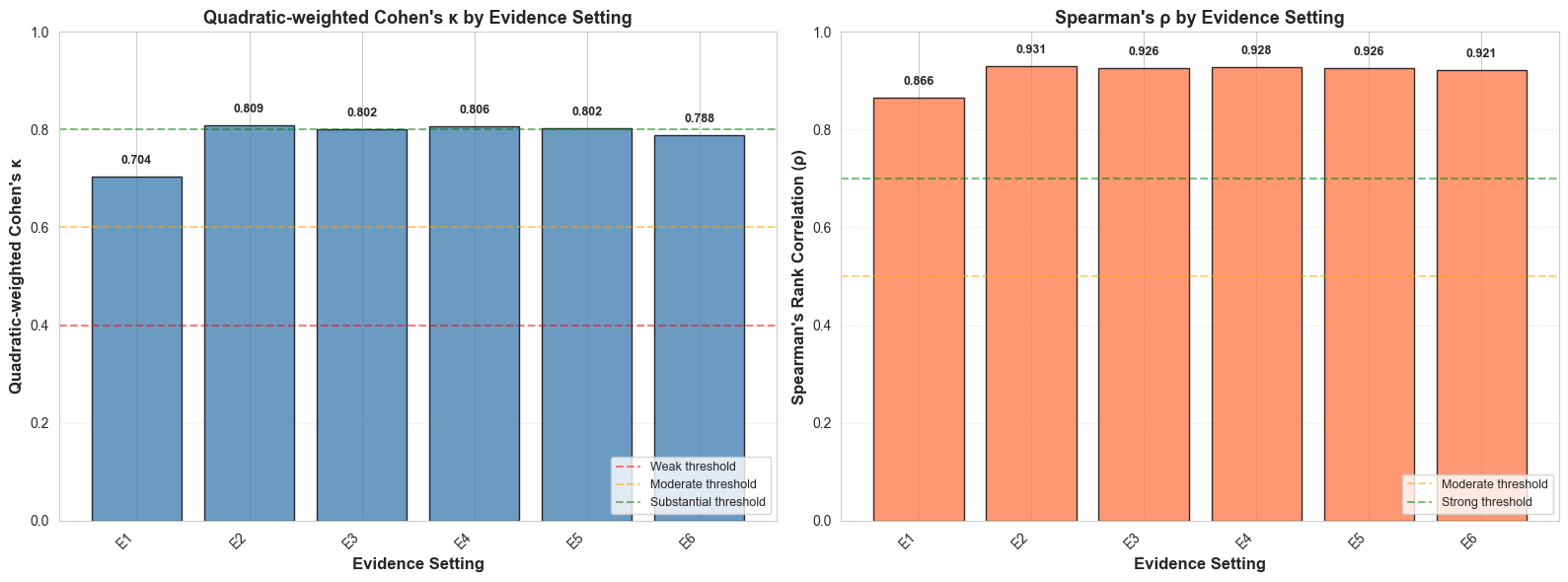}
\Description{Two panel chart showing quadratic-weighted Cohen's kappa and Spearman's rank correlation between the Qwen3-235B and Claude Haiku 4.5 judges across evidence protocols E1 through E6.}
\caption{Inter-judge agreement between Qwen3-235B and Claude Haiku 4.5 by evidence protocol.
Left: quadratic-weighted Cohen’s $\kappa$.
Right: Spearman’s rank correlation $\rho$.}
\label{fig:judge_agreement}
\end{figure}

\paragraph{Inter-judge agreement.}
Across all 15{,}600 paired evaluations, Qwen3-235B and Claude Haiku 4.5 show substantial ordinal agreement ($\kappa = 0.78$) and very strong rank correlation ($\rho = 0.94$, $p < 10^{-6}$; Figure~\ref{fig:judge_agreement}) \cite{cohen1968weighted, landis1977measurement, spearman1904proof}.
The judges exhibit a systematic calibration offset in absolute scoring (Qwen3-235B assigns higher mean scores), but relative rankings are highly consistent across all evidence settings, including noisier conditions.
Observed performance trends are therefore robust to judge choice.

\paragraph{SME judge calibration.}
We separately calibrate automated scores on a topology-balanced panel of 360 high-ranked retained QA pairs (20 per partition), distinct from the random QA-quality audit and intended to assess score concordance rather than benchmark-wide QA quality. A financial SME, blind to automated scores, graded GPT-OSS-120B responses (0--10) under E1, E2, and E6 (1,080 instances), representing KG-linked provenance, page-window evidence, and semantic retrieval. SME scores span the full 1--10 range and correlate strongly with both automated judges across these protocols (Spearman $\rho = 0.72$--$0.83$, $p < 0.001$), supporting their ranking reliability on this calibration panel.

\subsection{Statistical Significance of Evidence Effects}
\label{sec:results_significance}

To verify that the gains from KG-linked provenance evidence are not attributable to chance, we test whether E1 significantly outperforms each alternative evidence protocol (E2--E6). For each model and judge, we apply paired one-sided Wilcoxon signed-rank tests\cite{wilcoxon1945individual} on QA-level correctness scores (E1 $>$ Ei), with Holm--Bonferroni correction over the five comparisons per model--judge pair. We additionally report Cliff’s $\Delta$ to quantify practical effect magnitude. Across all model--judge combinations, E1 outperforms E2--E6 with statistically significant differences after correction (all $p_{\text{Holm}}<10^{-6}$) \cite{holm1979simple} and consistent positive effect sizes (median $\Delta \approx 0.29$).

\section{Conclusion}
\label{sec:conclusion}

FinMultiHopQA provides a large-scale, financially relevant, and diverse multi-hop QA benchmark, constructed via a controlled, scalable methodology (KG-driven motifs, GICS-conditioned sampling, and multi-tier validation), spanning intra-document, inter-year, and cross-company regimes with audited, provenance-linked evidence. Building on this construction, we show that reasoning models' judged correctness is sensitive to how evidence is structured: KG-linked provenance (E1) yields a \textbf{+20\%} correctness lift over less structured baselines while using fewer tokens, an ordering that persists across open-weight, frontier closed-source, and finance-specialized models alike. Larger models raise the correctness ceiling but do not change this ordering, indicating that reasoning capacity alone does not fully offset poorly structured retrieval. Beyond accuracy and efficiency, provenance-linked evidence gives auditors and compliance reviewers a traceable basis for verifying how an answer was derived from source disclosures.

\section{Limitations \& Future Work}
\label{sec:limitations}

FinMultiHopQA inherits the coverage limits of FinReflectKG, including gaps in entity extraction and relation types, and currently covers only S\&P~100 annual filings from 2022--2024; benchmark difficulty and model rankings may therefore shift across issuer populations, filing types, and reporting periods. Questions and reference answers are LLM-generated. Although multi-stage filtering and a two-SME audit of 360 topology-stratified pairs found high quality, the audit does not cover the full benchmark, so residual generation biases or localized answer errors may remain. Correctness is measured by LLM judges rather than human annotators at scale; we mitigate this through agreement between two LLM judges, SME calibration, and paired significance testing. Qwen3-235B serves as both a QA generator and an LLM judge, inducing a self/family-preference bias; the independent Claude Haiku 4.5 judge and the SME audit are both designed to guard against exactly this, and thus we read the results primarily as relative comparisons across models and protocols. Finally, the controlled evidence protocols are designed to isolate evidence-structure sensitivity, not to benchmark retrieval methods, and so do not represent deployed retrieval systems exhaustively. Future work will compare against end-to-end KG and non-KG retrieval baselines and isolate localization from graph structure through matched-evidence ablations. Reducing self-referential scoring further -- an independent rubric validator during QA construction, an expanded judge panel for evaluation, and a larger-scale SME audit -- is also a priority.

\bibliography{references}

\end{document}